# Multifunctional imaging enabled by optical bound states in the continuum with broken symmetry

Jiale Chen,[1, †] Zhao-Xian Chen,[2, †] Jun-Long Kou[1, 3, *] and Yan-Qing Lu[2, *]

[1]*School of Electronic Science and Engineering, Nanjing University, Nanjing 210093, P. R. China*
[2]*National Laboratory of Solid-State Microstructures and College of Engineering and Applied Sciences, Nanjing University, Nanjing 210093, P. R. China*
[3]*School of Integrated Circuits, Nanjing University, Suzhou 215163, P. R. China*
[†]*These authors contributed equally to this work.*
[*]*Correspondence should be addressed to jlkou@nju.edu.cn and yqlu@nju.edu.cn*

**Abstract:** For photonic crystal slab (PCS) structures, bound states in the continuum (BICs) and circularly polarized states (dubbed C-points) are important topological polarization singularities in momentum-space and have attracted burgeoning attention due to their novel topological and optical properties. In our work, the evolution of polarization singularities from BICs to C-points is achieved by breaking the in-plane $C_2$ symmetry of a PCS structure of a square lattice with $C_{4v}$ symmetry. Correspondingly, a BIC is split into two C-points with opposite chirality, incurring distinct optical transmission responses with the incidence of right or left circular polarization (RCP or LCP). Harnessing such chirality selectivity of the C-points, we propose a multifunctional imaging system by integrating the designed PCS into a conventional 4-*f* imaging system, to realize both the edge imaging and conventional bright-field imaging, determined by the circular polarization state of the light source. In addition to multifunctional imaging, our system also provides a vivid picture about the evolution of the PCS platforms' singularities.

## 1. Introduction

Light beams contain multidimensional information, such as wavelength, amplitude, phase, polarization, and orbital angular momentum [1]. Among them, the polarization properties of light waves have attracted enormous interest and provide a significant degree of freedom for optical manipulation with broadened applications in optical microscopy, multidimensional perceptions [2], optical communications [3], and so on. Recently, the PCS has been a popular platform for polarization-driven applications for its macroscopic size, ease of fabrication and simulations [4], and most importantly, its capability to realize all polarized states for full coverage on the Poincaré sphere [5]. Besides, the complicated far-field polarization states with different in-plane wave vectors $k_{//} = (k_x, k_y)$ in PCS structures can also be mapped onto the momentum space (k-space) and form the so-called polarization map, providing a deeper understanding of light from a topological perspective [6]. On the polarization map, there are several kinds of polarization singularities, such as centers of polarization vortices (V-points), lines of linear polarized states (L-lines), and C-points [7]. Those polarization singularities usually have unique physical properties potentially beneficial for various applications in modulating the radiation, polarization, phase, and transmission behaviors [8].

In the perspective of topological optics, singularities in the polarization map can be manipulated by symmetry operation to produce rich polarization field configurations in the momentum space [9]. Usually, symmetry-protected (SP) BICs emerge at the center of the Brillion zone center, namely the Γ point, for PCSs with $C_{4v}$ symmetry [5] or $C_{6v}$ symmetry [10] and appear as V-points in the polarization map. When perturbation is applied (e.g., broken symmetry) to introduce multipolar moments parallel to the sample plane [11], polarization fields in the momentum space change interestingly, accompanied by the destruction of SP-BICs. For example, the generation of six C-points from one V-point was numerically and experimentally verified with the symmetry of the system broken from $C_{6v}$ to $C_{3v}$ [10]; two-paired C-points were generated from one V-point with broken symmetry from $C_{4v}$ to $C_{2v}$ [5]. The evolutions of BICs and C-points are governed by the conservation law of topological charges [12]. Specifically, the V-point with integer charges splits into two C-points with half-integer charges. Accordingly, the nonradiative bound state changes into a resonance that can be excited by external light. The polarization singularities serve as a bridge that connects the radiation characteristics with the system's symmetry and have led to numerous applications in different areas. For instance, BICs (or V-points) in PCS platforms have been intensively investigated for a great deal of applications, such as BIC-based lasers [13, 14], enhanced nonlinear effects [15, 16], enhanced light-matter interactions [17, 18], and BIC-based sensing [19, 20], optical analog computing [21], etc. However, there are limited applications and devices enabled by the C-points except for a few demonstrations on chiral emission [22, 23], unidirectional guided resonances [24], arbitrary polarization conversion via PCSs [25], and PCS-induced polarization-dependent lateral beam shifts [26]. Recent work [5, 23] mentioned the chiral selectivity of C-points, while the underlying physics and great potential in applications of C-points is still far from being well explored. Thus, we propose a multifunctional imaging system based on C-points in the PCS, to promote the understanding of C-points and broaden the applications of C-points in the field of optical image processing.

In this Letter, we show that two-paired C-points with opposite chirality can emerge from the decomposition of a BIC in the PCS with broken symmetry. Due to the chirality selectivity of C-points in the proposed PCS device, two sets of optical transfer functions (OTFs) around the C-points are theoretically verified and are proved to be switchable under RCP or LCP incidence. We further demonstrate that our multifunctional imaging system, by integrating the PCS into a conventional 4-$f$ system [27], is capable of providing two different functions: bright-field imaging and edge-enhanced imaging [21, 28-30]. Notably, the proposed PCS operates in the transmission mode and works in the momentum space, providing nonlocal OTFs [31] for the two functions without requiring any collimation elements, making it more compatible with image processing applications. It is also worthy to mention that this nanophotonic system shows decent edge detection effects in both horizontal and vertical directions with wavelength-scale high resolution (1.10 μm). In all cases, our work provides a new path for the on-purpose design of different topological polarization singularities and broadens the applications of C-points in the image processing area.

## 2. Design and Principles

We start our discussions on the design of the proposed free-standing hole-type PCS, which supports transformable topological polarization singularities via symmetry operations [9]. For example, BICs in PCSs are known to be feasible for providing C-points, which are of great significance in solving problems of chiral photonics [32, 33]. As schematically shown in Fig. 1a, by breaking the in-plane $C_2$ symmetry of a PCS with $C_{4v}$ symmetry, two paired C-points with opposite chirality emerge from the destruction of the BIC. With this approach, we can follow a more intuitive way to design the desired PCS, compared with other designing approaches based on counterintuitive optimization methods [34]. The PCS consists of a square lattice of $Si_3N_4$ with a refraction index of $n = 2.02$, and a period of $a = 450$ nm. The thickness of the slab $d$ and the geometry (shape and dimension) of air holes offer adequate degrees of freedom to modulate the desired functions for the PCS device. According to temporal coupled mode theory [35, 36], due to the Fano interference between the resonance-assisted and background transmissions [37-40], namely the indirect and direct pathways in the photonic crystals, the total transmission usually exhibits asymmetric Fano-like profile, which is adverse for edge detection. However, transmission spectra with a Lorentz line shape are well suited for the perfect edge detection effect [41] needed in our multifunctional imaging system. To realize such a symmetric and Lorentz-like transmission profile around the C-points, here we set $d =$ 200 nm and optimize the duty ratio [36] (defined as the ratio of the air hole's area to the area of square unit cell) to $r = 0.30$ based on effective medium theory [42] to get the background transmission as $t_d = 1$, rendering a Lorentz-like profile at the resonant frequency [37]. Due to the chirality selectivity enabled by the C-points, the Lorentz-like profile turns into full transmission spectra when switching the helicity of incident light. Thus, another function, i.e., bright field imaging, can also be realized in our multifunctional imaging system.

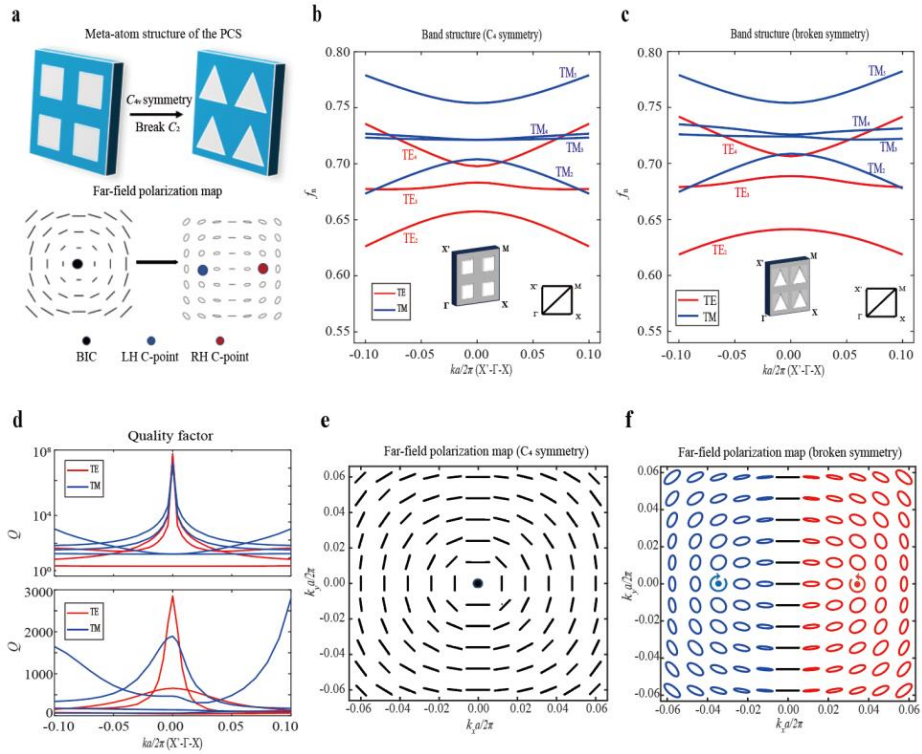

**Fig. 1**. Schematic for breaking the BIC point into two-paired C-points. **(a)** By breaking the in-plane $C_2$ symmetry (changing square air holes to isosceles triangle air holes, a SP-BIC splits into two C-points with opposite chirality. **(b-c)** The simulated bands of TE and TM modes of the PCS following the $X'$-$\Gamma$-$X$ direction, the inset of which shows the schematic of the proposed PCS and the first Brillion zone, respectively. **(d)** The $Q$ factors of bands of the PCS before (upper panel) and after (lower panel) symmetry breaking. **(e-f)** The extracted polarization field map of band $TE_2$ before (**e**) and after (**f**) symmetry breaking.

In order to determine the structural parameters of our PCS device and the physical peculiarity of different polarization singularities, eigenmodes analysis has been applied to simulate the transverse electric (TE) and transverse magnetic (TM) band structures and corresponding Quality factors $Q$s of the PCS with or without $C_{4v}$ symmetry. For the PCS with $C_{4v}$ symmetry, the air holes are square (see inset of Fig. 1b) with side length $L_0 = 247$ nm (the duty ratio $r =$ 0.30). At the $\Gamma$ point, the SP-BICs are identified as nonradiative states with diverging $Q$s, coexisting with extended modes within the light cone. According to the band structure in Fig. 1b and the $Q$ factors in the upper panel of Fig. 1d, we can clearly identify four SP-BICs located in four non-degenerate bands $TE_2$, $TE_3$, $TM_2$, $TM_5$ in the frequency range of interest: 0.62-0.80$f_0$ (namely 562-726 nm in the visible spectrum, where $f_0 = c/a = 666.7$ THz, $c$ is the speed of light in the vacuum). It is worth noting that the BICs cannot be excited by the normal incidence of any polarization and gradually evolve into leaky resonances at oblique incidence [32]. For the PCS with broken $C_2$ symmetry, the square air holes are changed to isosceles triangles with the area unchanged (see inset of Fig. 1c). Though there is a subtle change in the band structure in Fig. 1c, the $Q$s in the lower panel of Fig. 1d decrease significantly, meaning the BICs (nonradiative states) evolve into resonances (radiative) after breaking the in-plane $C_2$ symmetry. Without loss of generality, we focus on the band $TE_2$ with the operation frequency near 0.6375$f_0$ in the following design and discussions.

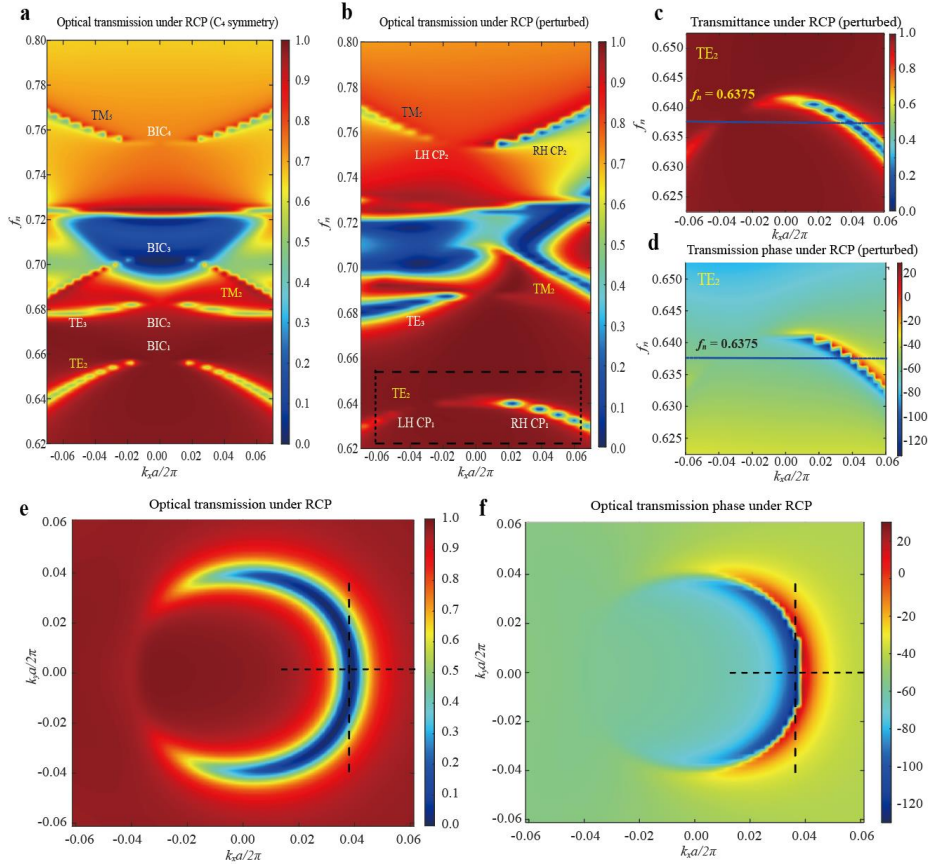

**Fig. 2**. Simulated angle-resolved transmission spectra and phase response under RCP incidence. **(a)** The simulated transmittance in the 2D parameter space ($k$, $f_n$) for the PCS with square air holes ($C_{4v}$ symmetry). Four bands ($TE_2$, $TE_3$, $TM_2$, $TM_5$) and four different SP-BICs are marked for the eye guide. **(b)** The simulated transmittance for the PCS with isosceles triangle air holes (broken symmetry). **(c-d)** The simulated optical transmission amplitude and phase response for $TE_2$ denoted with the dashed box in (**b**). **(e-f)** The optical transmission amplitude and phase response in the momentum-space ($k_x$, $k_y$) for $f_n = 0.6375$ marked as the horizontal lines in **c-d**. The two dotted crosses centered at the RH C-point indicate the field of view (Fig. 3) for the designed PCS.

To identify the radiation characteristics of the two PCS structures, we further calculate the distribution of far-field polarization, to map each intrinsic Bloch mode on the band $TE_2$ with different in-plane wave vectors in momentum-space (Figs. 1e-f). In Fig. 1e, the polarization map of the PCS with $C_{4v}$ symmetry clearly shows that far-field polarization states form a vortex around Γ, where the polarization is ill-defined at the V-point (denoted as the black dot). Notably, the corresponding SP-BIC can be characterized with a topological charge of -1, which is defined as the winding number of the vortex around the singularity [43]. Importantly, the polarization states of radiation from two-dimensional (2D) periodic structures are theoretically proved to be generally elliptical. For a lossless PCS system with in-plane $C_2$ symmetry and reciprocity, the polarization ellipticity of each radiative Bloch mode is often close to 0, making the whole polarization field around the V-point (Fig. 1e) close to linear polarization [43, 44], which explains why the PCS with $C_{4v}$ symmetry only supports linear polarization selectivity. When the in-plane $C_2$ symmetry is broken, as shown in Fig. 1f, the linear polarized states change to multiple elliptically polarized states, a L-line, a right-handed (RH) C-point, and a left-handed (LH) C-point. Note that the two C-points (denoted as the red and blue dots in Fig. 1a) are mirror-symmetric to each other about the $k_x = 0$ line owing to the preserved left-right mirror symmetry, while the L-lines (denoted as the black lines) lie on the mirror axis. Here, $k_x$ represents the normalized wavevector with respect to $k_0 = 2\pi/a$. The two C-points emerge at two specific $k$ points ($k_{CL(CR)} = \pm 0.039k_0$, where the "±" sign corresponds to the RH and LH C-point) near the Γ point in the horizontal direction. Besides, the conservation law of topological charges still holds for the evolution of different polarization singularities [43], the SP-BIC at Γ with a topological charge of -1 is split into two C-points with the same topological charges of -1/2 with opposite chirality. Note that the LH (or RH) C-point can be excited by LCP (or RCP) incidence while it cannot be accessed with opposite incidence, even though the $Q$, as shown in the lower panel of Fig. 1d, is finite with a magnitude of hundreds. Such chirality selectivity provides a new degree of freedom (the handedness of incidence) to modulate the far-field radiation and the in-plane transmission effects.

In order to get a better grasp of the different optical transmission responses of the PCS with or without broken symmetry, we simulate the angle-resolved transmission spectra and the transmission phase responses in the 2D parameter space ($k_x$, $f_n$) under circularly polarized incidence (we only present the results under RCP incidence for symmetry consideration, and the parameter $f_n$ represents the normalized frequency with respect to $f_0 = c/a$). For unperturbed PCS with RCP incidence, as shown in Fig. 2a, there are only four unexcited points on the bands of interest at the Γ point in the symmetric angle-resolved transmission spectra, indicating four SP-BICs (or vortex polarization singularities). According to the symmetry, the spectra do not change when the incidence switches to LCP, meaning the BICs cannot be excited by incident light of any circular polarization. However, for the case without $C_2$ symmetry, the transmission spectra under RCP incidence (Fig. 2b) become asymmetric due to the different exciting responses of two C-points. The asymmetrically vanishing regions on these bands indicate the corresponding C-points, which means that the incident light cannot excite these Bloch modes; that is, the far-field radiation polarization of these states is orthogonal to the polarization of the incident light and is circular. The mode with LCP (RCP) polarization eigenstate will not respond to the RCP (LCP) excitation and turn out to be a diminished point among the transmittance spectra. Specially, there is an unexcited region on the left side ($k_x < 0$, around $k_{CL} = -0.039k_0$ for LH C-point) of bands TE$_2$ while a continuous deep valley (transmittance undergoes a 1→ 0 →1 process) on the right side ($k_x > 0$, around $k_{CR} = +0.039k_0$ for RH C-point). The positions and handedness of the C-points are in accordance with the simulated polarization map (Fig. 1f). From such asymmetric transmission response, we can conclude that there is strict one-to-one correspondence between the C-points and the incidence, viz, the RCP (LCP) incidence can only excite the RH (LH) C-point, but not the opposite. In other words, the incidence with opposite chirality completely decouples with the nanostructure at the C-points. Next, we focus on the desired optical band (TE$_2$) and calculate its transmission responses. (Figs. 2c-d). The background transmission satisfies $t_d = 1$ for $f_n = 0.6375$, with which we can get a significant jump in the transmittance amplitude and an abrupt change in the phase spectra with the excitation of RCP incidence. At the RH C-point, LCP incidence totally decouples with the PCS, leading to unitary transmission through the structure, while the RCP incidence couples with the PCS and leads to destructive interference of the transmitted light, showing perfect reflection from the PCS. We further fixed the incident frequency at $0.6375f_0$ and simulated the optical transmission responses in the momentum space ($k_x$, $k_y$) (Figs. 2e-f) to better illustrate the peculiar properties of C-points ($k_y$ represents the normalized wavevector in the y direction). Under RCP incidence, the right side of Fig. 2e ($k_x > 0$) shows a significant change, from complete transmission (background $t_d = 1$) to zero transmission; however, the left side ($k_x < 0$) barely changes from the background. In particular, the transmission spectra with RCP incidence show an asymmetric crescent pattern centered around the RH C-point (Obviously, there is also a crescent pattern centered around the LH C-point under LCP incidence for symmetry consideration). It is worth noting that the simulated phase spectra of transmission coefficients (Fig. 2f) also show the same crescent pattern with a significant phase jump around the RH C-point. Such a peculiar asymmetric transmission response for LCP (or RCP) incidence reveals solid evidence of the chirality selectivity of the C-points. Therefore, C-points in PCS platforms have been demonstrated with its chirality selectivity, which makes this PCS system potentially suited for various chiro-optical applications, such as vortex beam generation [45], modulation in radiation characteristics and transmission behaviors [23], laying the foundation of our proposed multifunctional imaging system. In the following, we will demonstrate that the designed PCS device can provide the desired OTFs for different functionalities that our nanophotonic compound system requires.

We utilize the distinct transmission responses of C-point with different circularly polarized incidences to design the OTFs and to realize multifunctional imaging, i.e., edge detection and bright-field imaging. Without loss of generality, we chose the RH C-point ($k_{CR} = +0.039k_0$) for the rest demonstrations. Firstly, we extracted data from the simulations in Fig. 2 and fitted the OTFs centered around the RH C-point at frequency $0.6375f_0$ in the band $TE_2$ with the range: $|(k_x - k_{CR})/k_0| \leq 0.02$, $|k_y/k_0| \leq 0.02$ (field of view). Figure 3a-b and Figs. 3d-e show the fitted transmittance and phase response in the range of interest for RCP incidence and LCP incidence, respectively. We further calculated the transmission spectra along horizontal ($k_y = 0$) and vertical ($k_x = 0$) directions (Figs. 3c and 3f) denoted by the dotted crosses in Fig. 3a and Fig. 3d. For RCP incidence, the transmittance undergoes a jump from 1 to 0 around the RH C-point (Fig. 3a), and there is also an abrupt jump in the transmission phase (Fig. 3b) (Note that the OTFs here are of complex values; the real part and imaginary part of the OTF with RCP incidence show a similar crescent pattern). One can see that the transmittance spectra undergo a perfect Lorentz line shape horizontally while changing relatively slowly in the vertical direction. The Lorentz line shape is known to be well-fitted for edge detection [41]; most light reflects from the PCS except for the edges in the horizontal direction. The vertical direction experiences a similar situation except for smaller light transmittance. Thus, the OTF in the vertical direction is also able to distinguish the vertical edges in the given field of view. By contrast, the transmittance with LCP incidence is close to unity (Fig. 3d) in the given field of view, and the phase changes continuously (Fig. 3e). Meanwhile, the transmittance approaches unitary for both directions (Fig. 3f), which means most light transmits through the PCS under LCP incidence. The distinct OTFs for RCP and LCP incidence on the proposed PCS platform have been theoretically verified with the capability to realize 2D edge detection operation and bright-field imaging in the transmission mode. Thus, we can design a multifunctional imaging system by integrating the PCS into a conventional 4-$f$ imaging system. As schematically shown in Fig. 3g, the PCS was directly placed in front of the object plane, and the edge detection (or bright-field imaging) function is clearly exhibited under RCP (or LCP) incidence.

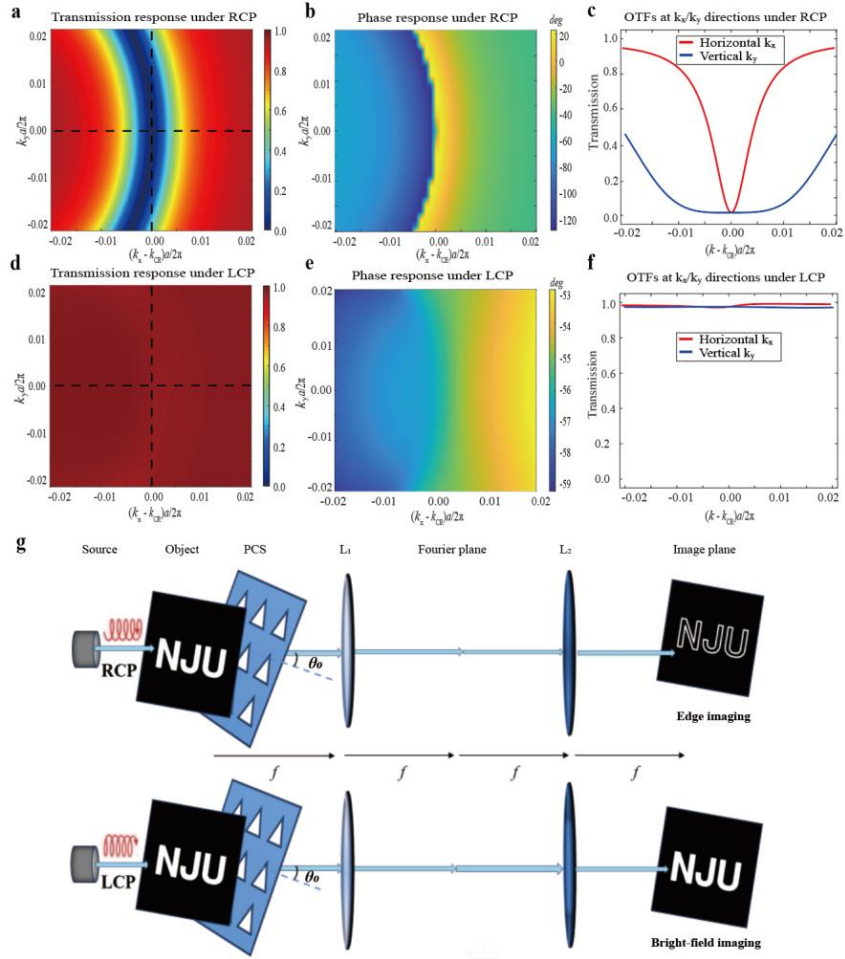

**Fig. 3**. The OTFs of the proposed PCS device around the RH C-point under RCP or LCP incidence, and the schematic of the designed multifunctional imaging system. The operation range of interest in the momentum space is $|(k_x - k_{CR})/k_0| \leq 0.02$ and $|k_y/k_0| \leq 0.02$. **(a-b)** The fitted transmittance (a) and transmission phase (b) under RCP incidence. **(c)** The transmittance at the positions marked by dotted lines in **a**. **(d-f)** The same as (**a-c**) but for LCP incidence. **(g)** Integrating the PCS into a conventional 4-f imaging system can realize edge (or bright field) imaging with RCP (or LCP) incidence.

## 3. Results and discussion

To numerically demonstrate that the device can be integrated into conventional imaging systems and verify its multifunctionality, we numerically simulate the Fresnel diffraction [46] and the OTFs of the PCS of the designed compound imaging system. As shown in Fig. 3g, the 4-$f$ system consists of two lenses ($L_1$ and $L_2$) of equal focal length $f$; thus, a "−1 magnification imaging" can be achieved at the image plane [47]. By directly placing the PCS behind the object plane in the 4-$f$ system, we can easily load the OTFs of the PCS into this compound nanophotonic system. Remarkably, the OTFs work at the momentum space owing to their nonlocal wavevector-dependent transfer functions [31, 48]. Therefore, the PCS is a nonlocal optical device, and we can get the same image at the output plane when the PCS is placed at the Fourier plane or other positions in the 4-$f$ system. (The imaging results when we place the PCS at the Fourier plane also show good performance of both edge-imaging and bright-field imaging). This position independence is a crucial advantage of our PCS device over most traditional optical elements that are local optical elements with position-dependent optical responses and need to be strictly aligned to the optic axis [48]. Furthermore, it is crucial that the PCS needs to be placed obliquely with a particular angle $\theta_0$ (Fig. 3g) with respect to the optic axis because the working range of OTFs is centered around the RH C-point (off-$\Gamma$) in momentum space, where $\theta_0 = \arcsin(k_{CR}/k_0) = 2.3$ deg. The target band is TE$_2$ (Fig. 2c), and the corresponding operation frequency is $f_w = 0.6375 f_0$ (705.4 nm, red light).

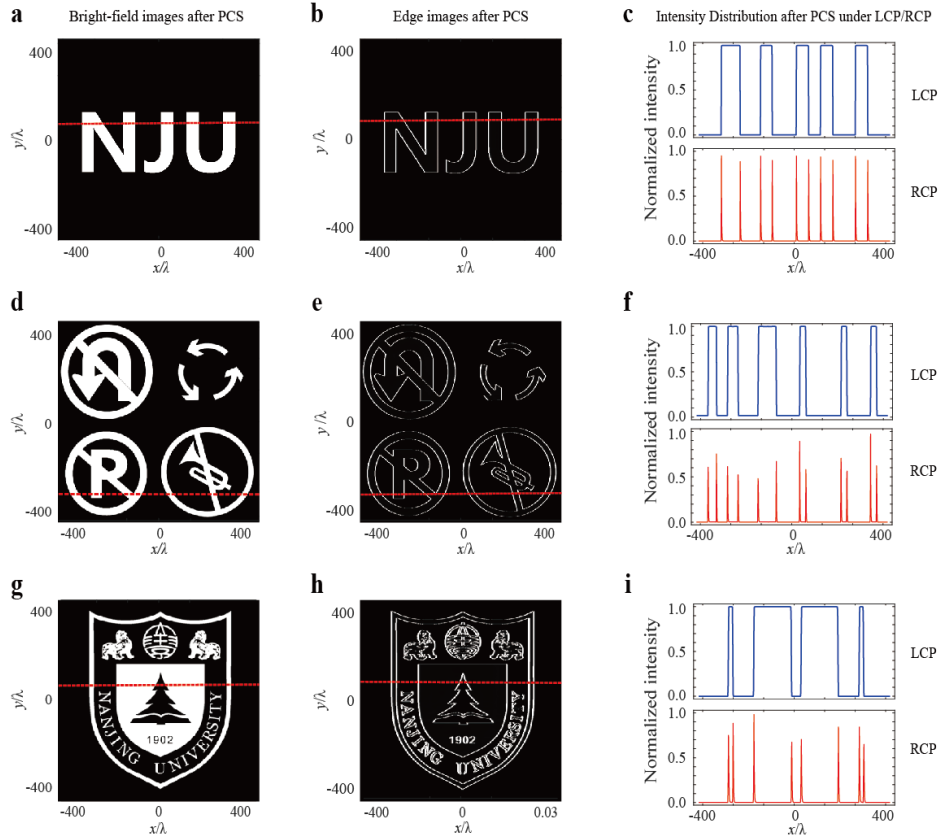

**Fig. 4**. Two different functions under RCP and LCP incidence of the proposed imaging system, edge-imaging and bright-field-imaging. **(a-b)** The normal bright-field images under LCP incidence and the edge-enhanced images under RCP incidence for Letters "NJU", respectively. **(c)** The measured intensity distribution of the bright-field images (under LCP incidence) and the edge-enhanced images (under RCP incidence) of the Letters "NJU". **(d-f)** The same as **(a-c)** but for the traffic signs. **(g-i)** The same as (a-c) but for the Logo of Nanjing University. Note that the dashed lines indicate the positions of our horizontal-cut measurements of intensities.

To evidently demonstrate the two different functions of the proposed imaging system, we performed theoretical simulations on three samples (Letters "NJU", traffic signs, and the logo of Nanjing University) in our imaging system. As shown in Fig. 3g, edge detection and bright-field imaging are switchable with different excitation modes (RCP or LCP). It is worthy pointing out that the original images of three cases are binarized for simplification. We recorded the output images for different cases, and the reversed images in the image plane were flipped back for convenient comparisons. Figure 4a, 4d, and 4g show the output images of three cases under LCP excitation; these bright-field images are of high image quality and no distortion and carry nearly all information of the input objects. When switching LCP to RCP, output images of high-contrast of the edges for three samples are presented as Figs. 4b, 4e, and 4g. It is remarkable that the edges of the letters, signs and logo are clearly revealed along both horizontal and vertical directions, which indicates a decent 2D edge detection effect with high resolution. Considering the transmittance through the PCS (different OTFs in two directions, Fig. 3c) discussed in the previous context, there is less light transmitting through the PCS in the vertical direction, so that the vertical edges are less obvious (but still distinguishable to some extent) than the horizontal ones. We further calculated the normalized intensity distribution in the horizontal direction for both edge-enhanced images and bright-field images in order to quantificationally examine the performance of our multifunctional imaging system. As shown in the intensity diagrams (Figs. 4c, 4f, 4i), where the upper panels show the amplitude of bright-field images, and the lower panels show the intensity distribution for edge-images. The red dashed lines in Fig. 4 indicate the positions of horizontal-cut measurements. It can also be seen that there are clearly visible separated sharp peaks located at every edge, and positions away from the edges are of nearly zero amplitude, which also supports the high-contrast imaging of the edges and evidently shows the perfect edge detection due to the Lorentz line shape OTF in the $k_x$ direction. By comparing the intensity distribution of the different cases, two functions of our multifunctional imaging system are more intuitively demonstrated. Note that we can get the minimum resolution of $1.56\lambda$ (1.10 μm, average widths at half-height of each sharp peak in the intensity diagrams of edge-enhanced images) to clearly distinguish all the edges, where $\lambda$ is the operation wavelength of the incident light, $\lambda = c/f_w = 0.706$ μm. Thus, we can conclude that bright-field imaging with high quality and edge imaging with high resolution (wavelength scale) have been successfully demonstrated, and the two different functions are switchable with RCP or LCP excitation.

One major advantage of the multifunctional imaging system is the nonlocality of the PCS device. This edge enhancement effect in our system is similar to dark-field imaging [49] but without the use of additional collimation components (i.e., a condenser) due to the nonlocal OTFs, which significantly reduces the system complexity. Besides, the PCS can be placed at an arbitrary position in the optical pathway in the 4-$f$ system owing to the nonlocality [31, 48]. Furthermore, the proposed multifunctional imaging system can operate over a relatively broad band due to the low-$Q$ resonance ($Q \sim 200$) away from the BIC state [50]. Figure 2c-d indicates that it is useful for edge discrimination across a broad bandwidth from $0.630f_0$ to $0.642f_0$, namely 700-714 nm. Additionally, the operation frequency of the PCS device is scaled as c/$a$, and it can be tuned accordingly by changing the period $a$ of the PCS based on different demands. We can also utilize the C-points at another band (i.e., TE$_3$ or TM$_5$) of the same structure. Even though we might get different topological configurations of C-points [5], we can achieve similar edge detection effects and chirality selectivity that can be utilized in our multifunctional imaging system.

## 4. Conclusion

In summary, we have proposed a novel approach to realize a multifunctional imaging system by utilizing the chirality of C-points of the proposed PCS structure. In our design, the two C-points of opposite chirality originate from the SP-BIC in the PCS with $C_{4v}$ symmetry when the in-plane $C_2$ symmetry is broken. Two sets of OTFs can be obtained from the peculiar asymmetric optical transmission response of C-points, forming the basis of our multifunctional imaging system. Then, our imaging system has been theoretically demonstrated to provide two different switchable functions: edge imaging under LCP incidence and bright-field imaging under RCP incidence. The proposed imaging system manifests the advantage of nonlocality, reduced complexity, high resolution (edge detection), broadband operation as well as the ability to implement two different OTFs. Not only does this compound nanophotonic system open new opportunities in applications such as biological imaging and computer vision, it also promotes the understanding and on-purpose design of V-points and C-points, and inspires new explorations in for novel phenomena and potential applications in radiation modulating, topological photonics, and image processing, which are worthy of further study.